\documentclass[prc,twocolumn,showpacs]{revtex4}

\usepackage{graphicx}
\usepackage{dcolumn}
\usepackage{bm}

\begin{document}

\title{$\beta$-decay properties of neutron-rich rare-earth isotopes}

\author{P. Sarriguren}
\email{p.sarriguren@csic.es}
\affiliation{
Instituto de Estructura de la Materia, IEM-CSIC, Serrano
123, E-28006 Madrid, Spain}

\date{\today}

\begin{abstract}

In this paper, $\beta$-decay properties of even-even neutron-rich isotopes in 
the rare-earth mass region are studied within a microscopic theoretical 
approach based on a proton-neutron quasiparticle random-phase approximation. 
The underlying mean field is constructed selfconsistently from a deformed 
Hartree-Fock calculation with Skyrme interactions and pairing correlations
to which particle-hole and particle-particle residual interactions are added.
Nuclei in this mass region participate in the astrophysical rapid neutron
capture process and are directly involved in the generation of the rare-earth 
peak in the isotopic abundance pattern centered at $A\simeq 160$. The energy 
distributions of the Gamow-Teller strength as well as the $\beta$-decay 
half-lives and the $\beta$-delayed neutron-emission probabilities are discussed 
and compared with the available experimental information and with calculations 
based on different approaches. 

\end{abstract}


\maketitle

\section{Introduction}

The astrophysical rapid neutron capture process ({\it r} process) is of fundamental 
importance to understand the nucleosynthesis of heavy neutron-rich nuclei and 
to account for their observed abundances \cite{bbhf,cowan91,arnould07,qian07}. 
It is believed that the process takes place in astrophysical explosive neutron-rich 
environments when the density of free neutrons is so high that the capture of 
neutrons by nuclei is the dominant mechanism and much faster than the competing 
$\beta$-decay. However, the identification of the possible physical sites for 
{\it r} process is still one major unsolved challenge. Neutron star mergers and 
neutrino-driven winds in core-collapse supernovae are two of the best suited 
scenarios (see \cite{arnould07,qian07} and references therein).

The path that the {\it r} process follows and the final pattern of isotopic 
abundances are the result of a complex network of reactions and decays competing 
to each other in changing conditions of densities and temperatures. Thus, to model 
the {\it r} process one needs reliable nuclear physics input for masses, neutron 
capture reaction rates, photodisintegration rates, $\beta$-decay half-lives 
($T_{1/2}$), and $\beta$-delayed neutron-emission probabilities ($P_n$) for a 
large amount of nuclei ranging from the valley of stability to the neutron drip 
line.

The sensitivity of the final abundance pattern to the various nuclear properties 
and astrophysical conditions has been largely studied (see for example 
Ref. \cite{mumpower12,mumpower15} and references therein). The main conclusion is 
that given the astrophysical conditions, the final abundance pattern is basically 
determined by the nuclear properties of nuclei along the {\it r}-process path, as 
well as on the properties of nuclei involved in the freeze-out at the last stages 
of the process leading nuclei back to stability. In particular, the peaks found 
in the {\it r}-process abundance pattern at $A\simeq 80$, $A\simeq 130$, and 
$A\simeq 195$ are associated with the relatively large $\beta$-decay half-lives 
in the waiting point nuclei with closed neutron shells $N=50$, $N=82$, and $N=126$, 
respectively. The matter accumulates at these points creating peaks in the 
observed abundances. At these points of extra stability the neutron capture 
becomes inhibited and the nucleosynthesis flow is highly influenced by the 
large $\beta$-decay half-lives that finally determine the timescale of the 
whole process.

Above $A\simeq 100$, in the region between the two main peaks at $A\simeq 130$ 
and $A\simeq 195$, there is also a smaller peak, which is away from closed neutron 
or proton shells. It appears at $A\simeq 160$ and is called the rare-earth peak.
It has been shown \cite{surnam97} that the rare-earth peak is generated during 
the last stages of the {\it r} process as nuclei decay to stability. The peak is 
mainly due to the combined effects of nuclear deformation and $\beta$-decay 
\cite{surnam97,mumpower14} and it is very sensitive to the nuclear properties 
of neutron-rich rare-earth isotopes of about 10-15 neutrons away from stability 
\cite{mumpower12}. 
On the other hand, it has also been argued that fission recycling with asymmetric 
fission fragment distributions may play a crucial role to understand the origin of 
the rare-earth peak \cite{goriely13}. This would emphasize the role of neutron 
star mergers as possible sites for the generation of the rare-earth peak. Thus, 
the characteristics of the rare-earth peak may offer unique insight into the 
late-time freeze-out behavior of the {\it r} process.

Unfortunately, it is difficult to measure the properties of nuclei far from 
the valley of stability involved in the {\it r} process because of the 
extremely low production yields in the laboratory. As a consequence, there is 
still little experimental information on the quantities needed for astrophysics.
Although much progress is being made to measure masses and half-lives 
\cite{jyvaskyla,pereira,nishimura11,lorusso15,quinn12} and there are promising 
perspectives for the next future in the new facilities at FAIR \cite{reifarth16}, 
RIBF-RIKEN \cite{wu16}, and FRIB-MSU \cite{wrede15}, the astrophysical simulations 
of the {\it r} process must still rely on extrapolations of the available data 
or on predictions from theoretical nuclear models. Obviously, these models must 
prove first their reliability by reproducing the available data, which in the 
case of decay properties means half-lives and Gamow-Teller strength distributions 
measured in the laboratory. 

In medium and heavy mass nuclei, the shell model needs very large configuration
spaces that exceed the present computation capabilities and the quasiparticle 
random-phase approximation (QRPA) has been shown to be well suited to describe
the properties of these nuclei and in particular, their decay properties.
QRPA calculations for neutron-rich nuclei have been performed in the past from 
spherical formalisms \cite{engel99,borzov08,niksic05}. However, although the 
mass regions of interest mentioned above surrounding the magic neutron numbers 
$N=50,82,126$ can be well described with such spherical formalisms, the 
{\it r}-process path crosses open-shell regions characterized by well deformed 
shapes. Thus, deformed QRPA calculations, such as those in Refs.
\cite{moller1,homma,hir2,hamamoto,sarri1,sarri2,sarri3,sarri4,fang13,ni14,yoshi13,peru14} 
have been developed. The sensitivity of the $\beta$-decay properties to nuclear 
deformation has been proven in the above references, showing that these properties 
may differ substantially from the spherical assumption and that it is necessary 
to include the nuclear deformation for reliable estimates. The sensitivity to 
deformation of the GT strength distributions has been exploited experimentally 
to be an additional source of information about the nuclear deformation, as it 
was shown in Ref. \cite{expnacher}. The effects of deformation on the decay 
properties of neutron-rich isotopes in medium-mass nuclei from Ge up to Pd have 
been studied in Refs. \cite{fang13,ni14,yoshi13,sarripere,sarri14,sarri15}, 
demonstrating the need of theoretical formalisms accounting for nuclear 
deformation.

The rare-earth region, including nuclei in the range between $82 < N < 126$ 
and $50 < Z < 82$, is also well known to accommodate good rotors. However, very 
few studies of the decay properties of rare-earth nuclei are available yet. 
There are global calculations combining a microscopic QRPA approach for the 
GT response with the statistical gross theory for the first-forbidden (FF) 
decay \cite{moller3}. There are also deformed QRPA calculations based on 
Woods-Saxon potentials and realistic CD-Bonn residual forces using the 
$G$-matrix \cite{fang16}. The finite-amplitude method has been used in Ref. 
\cite{mustonen16} to obtain a global description of the decay properties within 
a Skyrme QRPA with axial symmetry. Finally, a spherical relativistic formalism 
has been used for a large-scale evaluation of the $\beta$-decay properties in 
{\it r}-process nuclei \cite{marketin16}.

In this work I investigate the nuclear structure related to the decay properties
in the region of neutron-rich rare-earth nuclei of interest for the {\it r} 
process \cite{mumpower12} that includes  Xe, Ba, Ce, Nd, Sm, Gd, and Dy isotopes 
with neutron numbers $86 < N < 126$. A theoretical formalism based on a deformed 
Skyrme HF+BCS+QRPA approach is used to calculate energy distributions of the 
Gamow-Teller strength for those isotopes and weighting properly the strength with 
phase-space factors, $\beta$-decay half-lives and $\beta$-delayed neutron-emission 
probabilities are calculated. The results are compared with the available 
experimental information on half-lives \cite{audi12} and with other theoretical 
calculations \cite{moller3,fang16,mustonen16,marketin16}. Thus, after testing the 
capability of the method to reproduce the measured half-lives, predictions are
made in 
more exotic nuclei including some of the isotopes that are planned to be measured 
in the next future \cite{reifarth16,wu16,wrede15}. Also important for astrophysics 
are the GT strength distributions because they contain the underlying nuclear 
structure. Given that the phase factors in a stellar medium at high densities and 
temperatures may be quite different from the temperature in the laboratory, the stellar 
half-lives may also differ substantially from the laboratory half-lives because 
of the thermal population of the decaying nuclei and because the electron 
distribution in the stellar plasma might block the $\beta$-particle emission 
\cite{langanke}.

The article is arranged as follows:
First, in Sec. \ref{sec2} I review the theoretical formalism used in this work. 
Then, I show in Sec. \ref{results} the results obtained for the selfconsistent
deformations in the neutron-rich isotopic chains of Xe, Ba, Ce, Nd, Sm, Gd, and
Dy, together with the energy distributions of the Gamow-Teller (GT) strength at
the corresponding equilibrium deformations. I also calculate $\beta$-decay 
half-lives and probabilities for $\beta$-delayed neutron emission, comparing 
the results with experimental data and with results from other calculations. 
Section IV summarizes the main conclusions.

\section{Theoretical Formalism}
\label{sec2}

The theoretical formalism used in this work to study the decay properties of 
neutron-rich isotopes has been already introduced elsewhere 
\cite{sarri1,sarri2,sarri3,sarri4} 
and I only sketch here the main points. The method starts from a selfconsistent 
Hartree-Fock (HF) mean field calculation with Skyrme interactions that includes 
pairing correlations and deformation. 
The calculations are performed with the force SLy4 \cite{sly4}, which is a 
suitable representative of the Skyrme forces. The fitting protocol of the
parameters includes nuclear properties of unstable nuclei. The sensitivity of
the decay properties to different choices of the underlying effective force 
is not critical and has been studied elsewhere 
\cite{sarri1,sarri2,sarri3,sarri4,sarri5}. 
The single-particle wave functions are expanded in terms of the eigenstates of 
an axially symmetric harmonic oscillator in cylindrical coordinates, using 
twelve major shells. Pairing between like nucleons is considered in the BCS 
approximation by solving the BCS equations after each HF iteration, using 
fixed gap parameters which are determined from the experimental odd-even mass 
differences when this information is available \cite{audi12}. When the masses 
are unknown the same pairing gaps as the closer isotopes measured are used. 
The method provides naturally the selfconsistent equilibrium deformations 
that minimize the energy and there is no need to introduce any extra parameter.

In a next step, a spin-isospin residual interaction is added to the mean field 
and treated in a deformed proton-neutron QRPA
\cite{hir2,sarri1,sarri2,moller1,moller3,homma}.
The residual interaction contains two components acting in the particle-hole 
($ph$) and in the particle-particle ($pp$) channels. The former can be derived 
consistently from the same Skyrme interaction used in the mean field 
calculation in terms of the second derivatives of the energy density functional 
with respect to the one-body densities. To simplify the calculation, the $ph$ 
residual force is expressed in a separable form \cite{sarri1,sarri2}. 
This is a repulsive interaction that redistributes the GT strength by reducing
and moving it to higher excitation energies and thus, shifting the GT resonance.
The coupling strength is usually taken to reproduce the location of the GT 
resonance \cite{sarri1,sarri2,homma}. Based on previous work, the value 
$\chi ^{ph}_{GT}=0.15$ MeV is taken here. The $pp$ part is also introduced as a 
separable force \cite{hir2}. It is attractive and moves the GT strength to lower 
excitation energies. Its coupling strength is usually fitted to reproduce the 
experimental half-lives \cite{homma}.  Based on previous calculations (see Refs. 
\cite{sarri091,sarri092} and references therein), the value  
$\kappa ^{pp}_{GT} = 0.03$ MeV is taken.
One could fit these two coupling constants to reproduce the measured half-lives 
of particular nuclei, but because the aim here is to test the ability of the
method to account for the decay properties in a global way, I use the same 
coupling strengths for all the nuclei considered in this work.

The suitability of the separable forces has been tested in Refs. 
\cite{yousef09,fang10,fang13}
by comparing the results obtained from QRPA using deformed Woods-Saxon 
potentials and realistic CD-Bonn residual forces with similar calculations with
separable forces. It has been shown that separable forces still keep the main 
characteristic of realistic residual forces, concluding that both approaches, 
realistic and separable, lead to similar results.

Introducing the proton-neutron QRPA phonon operator that creates a GT excitation 
in an even-even nuclei as 

\begin{equation}
\Gamma _{\omega _{K}}^{+}=\sum_{\pi\nu}\left[ X_{\pi\nu}^{\omega _{K}}
\alpha _{\nu}^{+}\alpha _{\bar{\pi}}^{+}+Y_{\pi\nu}^{\omega _{K}}
\alpha _{\bar{\nu}} \alpha _{\pi}\right]\, ,  \label{phon}
\end{equation}
the intrinsic excited states $\left| \omega _K \right\rangle $ are given by

\begin{equation}
\left| \omega _K \right\rangle = \Gamma ^+ _{\omega _{K}} \left| 0 \right\rangle \, ,
\end{equation}
where the QRPA vacuum $\left| 0\right\rangle $ satisfies
\begin{equation}
\Gamma _{\omega _{K}} \left| 0 \right\rangle =0 \, .
\end{equation}
$\alpha ^{+}$ and $\alpha $ are quasiparticle creation and annihilation operators, 
respectively. $\omega _{K}$ labels the QRPA excitation energy and 
$X_{\pi\nu}^{\omega _{K}},Y_{\pi\nu}^{\omega _{K}}$ are the forward and backward amplitudes, 
respectively. In the intrinsic frame, the allowed GT transition amplitudes 
connecting the QRPA ground state to one-phonon states are given by

\begin{equation}
\left\langle \omega _K | \sigma _K t^{-} | 0 \right\rangle =
\sum_{\pi\nu}\left( q_{\pi\nu}X_{\pi
\nu}^{\omega _{K}}+ \tilde{q}_{\pi\nu}Y_{\pi\nu}^{\omega _{K}}
\right) ,
\label{intrinsic}
\end{equation}
with
\begin{equation}
\tilde{q}_{\pi\nu}=u_{\nu}v_{\pi}\Sigma _{K}^{\nu\pi },\ \ \
q_{\pi\nu}=v_{\nu}u_{\pi}\Sigma _{K}^{\nu\pi},
\label{qs}
\end{equation}
$v'$s are occupation amplitudes ($u^2=1-v^2$) and $\Sigma _{K}^{\nu\pi}$ spin matrix 
elements connecting neutron and proton states with spin operators
\begin{equation}
\Sigma _{K}^{\nu\pi}=\left\langle \nu\left| \sigma _{K}\right|
\pi\right\rangle \, .
\end{equation}

Expressing the initial and final states in the laboratory frame in terms of the 
intrinsic states using the Bohr-Mottelson factorization \cite{bm}, the GT strength 
$B_{\omega}(GT^-)$ for a transition $I_iK_i (0^+0) \rightarrow I_fK_f (1^+K)$ is 
obtained in terms of the intrinsic amplitudes in Eq. (\ref{intrinsic}) as
\begin{eqnarray}
B_{\omega}(GT^- )& =& \sum_{\omega_{K}} \left[ \left\langle \omega_{K=0}
\left| \sigma_0t^- \right| 0 \right\rangle ^2 \delta (\omega_{K=0}-
\omega ) \right.  \nonumber  \\
&& \left. + 2 \left\langle \omega_{K=1} \left| \sigma_1t^- \right|
0 \right\rangle ^2 \delta (\omega_{K=1}-\omega ) \right] \, ,
\label{bgt}
\end{eqnarray}
in $[g_A^2/4\pi]$ units. 
 
The GT strength distributions will be shown later as a function of the excitation 
energy $E_{ex}$ with respect to the ground state of the odd-odd daughter nucleus, 
$E_{ex} = \omega_{QRPA} - E_0$, obtained by subtracting the lowest two-quasiparticle 
energy $E_0$ from the calculated $\omega$ energy in the QRPA calculation.

$\beta$-decay half-lives are calculated by summing all the allowed transition 
strengths to states in the daughter nucleus with excitation energies lying below 
the corresponding $Q$-energy, and weighted with the phase space factors 
$f(Z,Q_{\beta}-E_{ex})$,

\begin{equation}
T_{1/2}^{-1}=\frac{\left( g_{A}/g_{V}\right) _{\rm eff} ^{2}}{D}
\sum_{0 < E_{ex} < Q_\beta}f\left( Z,Q_{\beta}-E_{ex} \right) B(GT,E_{ex}) \, ,
 \label{t12}
\end{equation}
with $D=6200$~s and $(g_A/g_V)_{\rm eff}=0.77(g_A/g_V)_{\rm free}$, where 0.77 is a 
standard quenching factor. The $Q_{\beta^-}$ energy is given by

\begin{eqnarray}
Q_{\beta^-}&=& M(A,Z)-M(A,Z+1)-m_e  \\
&=& BE(A,Z) -BE(A,Z+1)+m_n-m_p-m_e \, , \nonumber
\end{eqnarray}
written in terms of the nuclear masses $M(A,Z)$ or nuclear binding energies 
$BE(A,Z)$ and the neutron ($m_n$), proton ($m_p$), and electron ($m_e$) masses.

The Fermi integral $f(Z,Q_{\beta}-E_{ex})$ is computed numerically for each value 
of the energy including screening and finite size effects, as explained in 
Ref. \cite{gove}. This function increases with the energy of the 
$\beta$-particle and therefore the strength located at low excitation energies 
contributes more importantly to the half-life.

The $\beta$-delayed neutron-emission probability is calculated as

\begin{equation}
P_n = \frac{ {\displaystyle \sum_{S_n < E_{ex} < Q_\beta}f\left( Z,Q_{\beta}-E_{ex}
\right) B(GT,E_{ex}) }}
{{\displaystyle \sum_{0 < E_{ex} < Q_\beta}f\left( Z,Q_{\beta}-E_{ex} \right)
B(GT,E_{ex})}}\, .
\label{pn}
\end{equation}
Thus,  $P_n$ corresponds to the probability of neutron emission with no distinction
between emission of one, two, or more neutrons.
The sum extends to all the excited sates in the daughter nucleus with excitation 
energies within the indicated ranges. $S_n$ is the one-neutron separation
energy in the daughter nucleus. In this expression it is assumed that all the 
decays to energies above $S_n$ in the daughter nucleus lead to delayed neutron 
emission and then, $\gamma$-decay from neutron unbound levels is neglected. 
Thus, the probability is always overestimated.

\section{Results and discussion}
\label{results}

\begin{figure*}[htb]
\centering
\includegraphics[width=90mm]{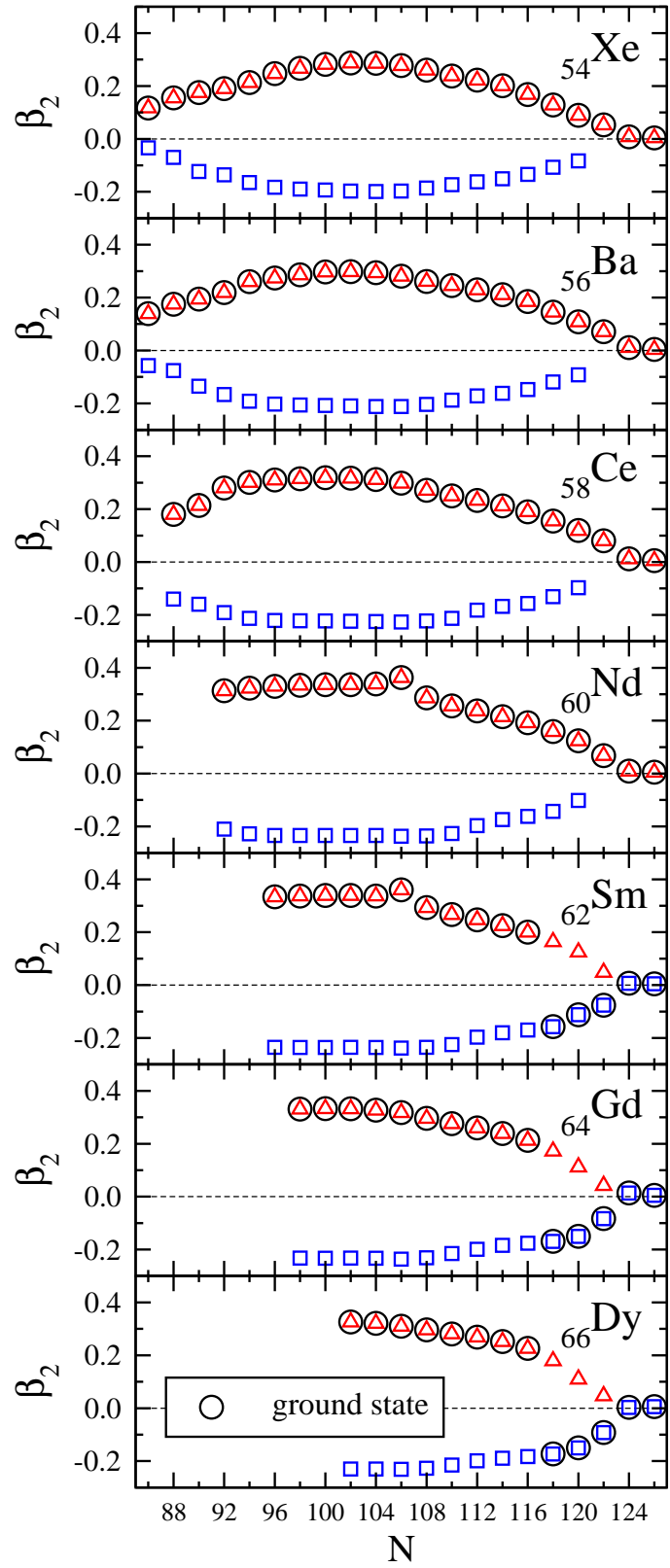}
\caption{(Color online) Isotopic evolution of the quadrupole deformation $\beta_2$ 
corresponding to the energy minima in neutron-rich Xe, Ba, Ce, Nd, Sm, Gd, and Dy
isotopes, obtained from SLy4. Ground state results for each isotope are encircled.}
\label{fig_beta}
\end{figure*}

Constrained HF+BCS calculations are performed to obtain energy curves
that show the energy as a function of the quadrupole deformation $\beta_2$.
Most of the energy curves in the isotopes studied in this work exhibit two 
local minima, prolate and oblate, separated by energy barriers that change 
with the neutron number. One can see in Fig. \ref{fig_beta} the isotopic 
evolution of the quadrupole deformations for the prolate and oblate equilibrium 
shapes plotted versus the neutron number $N$. The results correspond to the unstable 
isotopes of the neutron-rich rare-earth nuclei considered in this work. The 
deformation of the ground state for each isotope is encircled. Prolate ground 
state shapes are obtained in all the cases except some of the heavier isotopes that 
become oblate, and finally spherical when approaching the closed shell at $N=126$.
About the mid-shell one finds the largest deformations, as well as the largest 
energy barriers between the two minima. A very similar trend is observed in the 
profile of the curves in all the isotopic chains, showing larger deformations 
around mid-shell ($N=104$) with values close to $\beta_2=0.35$ in the prolate 
sector and close to  $\beta_2=-0.25$ in the oblate one. The quadrupole deformations 
become smaller as one approaches the closed shell isotopes with $N=82$ and $N=126$. 
The deformations obtained in this work are in good agreement with those obtained 
with the Gogny-D1S energy density functional that are available in Ref. \cite{gogny}.
The shape transitions from prolate to oblate shapes is predicted to occur in this 
work between $N=116$ and $N=118$ in Sm, Gd, and Dy isotopes. This is in 
qualitative agreement with the results from Gogny-D1S, where the transitions
are predicted at $N=116$ in Sm and Gd and at $N=114$ in Dy. Transitions at 
$N=118$ in Nd and at $N=120$ in Ce are also predicted with Gogny-D1S.
It is also worth comparing the above results with those from modern global
mass models, such as the semi-empirical nuclear mass formula based on 
macroscopic-microscopic methods \cite{minliu10}. This mass formula predicts a
shape transition from prolate to oblate at $N=118,\ 122,\ 122,\ 120,\ 118,\ 118,\ 116$ 
in Xe, Ba, Ce, Nd, Sm, Gd, and Dy, respectively. Thus, the transition takes place
somewhat earlier in the lighter nuclei Xe, Ba, Ce, and Nd, somewhat later in 
Sm and Gd, and at the same isotope $N=116$ in Dy.
Nevertheless, the important point is the shape-change tendency taking place before 
the closed shell at $N=126$, which is a general feature of the isotopic shape
evolution. The exact isotope where the transition takes place is not very
relevant because the shape transition region is naturally characterized by 
very close energies in the prolate and oblate sector competing to each other
to be ground states. Thus, small details of the calculations can change the 
energies and shift the ground state from one shape to another. 
Another odd effect observed in  Fig. \ref{fig_beta} is the presence of a 
little kink in the quadrupole deformations at $N=106$ in Nd and Sm 
isotopes. This kink is related to a subtle effect that appears at mid-shell 
in these nuclei, where the tendency changes and the equilibrium deformation 
starts decreasing with increasing neutron number. 
Looking in detail into the energy-deformation curves obtained from constrained 
HF+BCS calculations, one finds out that this behavior is related to the 
topology of the curves in the prolate minimum. Whereas the minimum up to 
$N=104$ is rather sharp, at $N=106$ it becomes shallower with the absolute 
minimum at a somewhat larger value $\beta_2=0.36$. At $N=108$ the prolate
minimum is also rather shallow, but slightly peaked at a lower $\beta_2=0.29$. 
After that, it continuously decreases with increasing $N$. It is hard to find 
a simple explanation for this rare behavior that is related to a very subtle 
competition between the single-particle energies and their crossing as a function 
of deformation in these isotopes. The effect might also depend on the particular 
version of the Skyrme interaction used (SLy4 in this case).

In the next figure (Fig.~\ref{fig_bgt}) the accumulated GT strength, that is, 
the GT strength contained up to a given excitation energy, is plotted as a 
function of the excitation energy of the daughter nucleus. A quenching factor 
0.77 has been included in the results. The figures cover the energy range below 
the $Q_\beta$ energy, which is the relevant region for the calculation of the 
half-lives. Only the results for the lighter unstable isotopes are shown because 
they offer better possibilities to be measured. Vertical arrows indicate the 
experimental $Q_\beta$ energies \cite{audi12}.

The energy distribution of the GT strength is fundamental to constrain the 
underlying nuclear structure involved in the calculation of the the half-lives. 
The decay rates in astrophysical scenarios may however, be different from the
decay rates
in the laboratory because the phase factors may be also different. Therefore, 
to describe properly the decay rates under extreme conditions of density and 
temperature, it is not sufficient to reproduce the half-lives in the 
laboratory. One needs, in addition, to have a reliable description of the GT 
strength distributions \cite{sarri091,sarri092}. The different profiles observed 
for the prolate an oblate nuclear shapes is a typical example of the sensitivity 
of the GT strength distribution to deformation. This sensitivity is translated 
into the $\beta$-decay half-lives. In the heavier isotopes Ce, 
Nd, Sm, and Gd, the oblate shapes generate more GT strength than the prolate 
ones at low excitation energies and as a result, the oblate shapes produce 
shorter half-lives.
Experimental information on these strength distributions will be very valuable
to constrain further the nuclear structure calculations.

\begin{figure*}[htb]
\centering
\includegraphics[width=90mm]{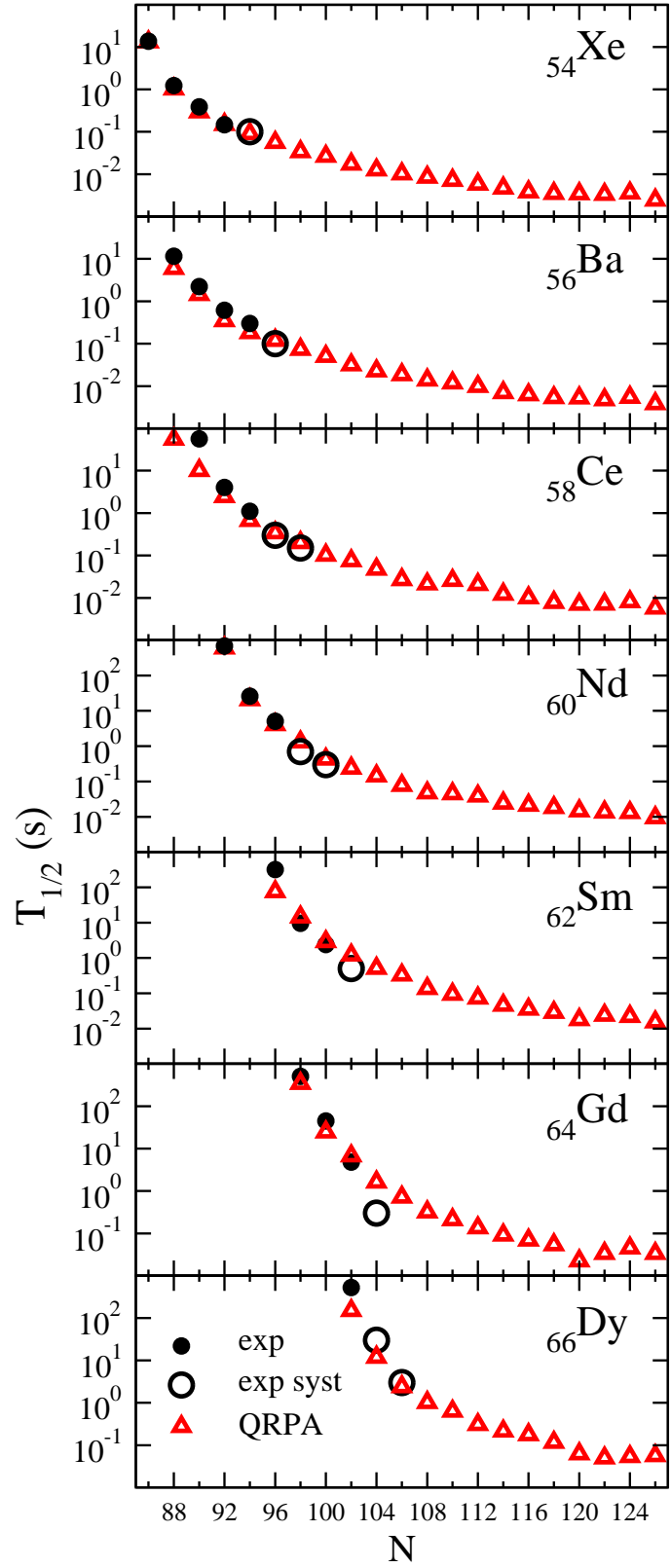}
\caption{(Color online ) Theoretical QRPA-SLy4 $\beta$-decay half-lives compared 
with experimental half-lives \cite{audi12} (open circles stand for experimental 
values from systematics) for neutron-rich Xe, Ba, Ce, Nd, Sm, Gd, and Dy isotopes.}
\label{fig_t12}
\end{figure*}

The calculation of the half-lives in Eq. (\ref{t12}) relies on the GT strength 
distribution and $Q_\beta$ values. In this work, values obtained from SLy4 
\cite{masssly4} are used. In Fig. \ref{fig_t12} the measured $\beta$-decay 
half-lives (solid dots), including the values extracted from systematics (open 
dots) \cite{audi12}, are compared with the theoretical results of this work 
for the various isotopic chains. In general, a very reasonable agreement with the 
few experimental measurements in this mass region is obtained.

\begin{figure*}[htb]
\centering
\includegraphics[width=80mm]{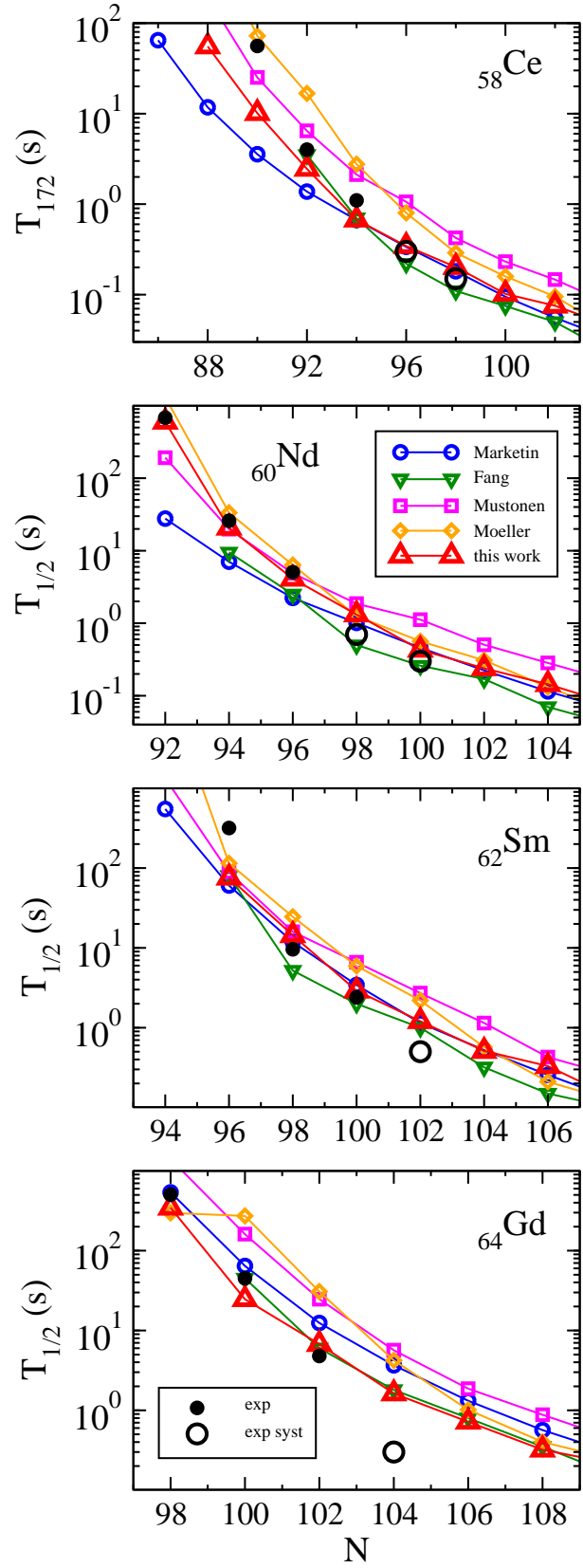}
\caption{(Color online ) Theoretical QRPA-SLy4 $\beta$-decay half-lives (this work)
compared with results from Marketin {\it et al.} \cite{marketin16}, Fang \cite{fang16},
Mustonen {\it et al.} \cite{mustonen16}, and Moeller {\it et al.} \cite{moller3}.
Experimental data are as in Fig. \ref{fig_t12}. }
\label{fig_t12_comp}
\end{figure*}

In Fig. \ref{fig_t12_comp} I compare the results for the half-lives obtained in 
this work with the results in Refs. \cite{moller3,fang16,mustonen16,marketin16} 
and with experiment. Results are shown for the Ce, Nd, Sm, and Gd isotopes 
around the region where the half-lives are measured. The calculations in 
Ref. \cite{moller3} combine a microscopic QRPA approach based on a Yukawa 
single-particle Hamiltonian and a separable residual interaction in the $ph$ 
channel for the GT response with the statistical gross theory for the 
first-forbidden decay \cite{moller3}. The $Q_{\beta}$ and neutron separation 
energies are taken from the masses calculated in the finite-range droplet-model 
(FRDM) \cite{frdm}. The calculations are done without any quenching of the 
axial-vector coupling constant $g_A$. The results in Ref. \cite{fang16} are 
obtained from deformed QRPA calculations based on Woods-Saxon potentials and 
realistic CD-Bonn residual forces using the $G$-matrix. The strength of the 
$pp$ residual interaction is renormalized by taking into account the Pauli 
exclusion principle that generalizes the usual quasiboson approximation. This 
procedure avoids using effective parametrizations of the the $pp$ force. The 
masses are also taken from the FRDM \cite{frdm} and a standard quenching is 
included. Global calculations of decay properties based on Skyrme QRPA for 
axially deformed even-even nuclei have been performed in Ref. \cite{mustonen16} 
within the finite-amplitude method. Allowed and FF transitions are both 
considered. The results in Ref. \cite{marketin16} correspond to spherical 
calculations within a relativistic formalism including FF transitions. In 
the last two cases the masses are calculated consistently, but whereas the 
quenching of $g_A$ is considered for both GT and FF transitions in 
Ref. \cite{marketin16}, it is only included for GT transitions in 
Ref. \cite{mustonen16}.

Figure \ref{fig_t12_comp} shows that the half-lives calculated in references 
\cite{moller3} and \cite{mustonen16} have a tendency to be above the half-lives 
calculated 
in references \cite{marketin16} and \cite{fang16}. The calculations in this work 
appear between the results of the other approaches, except in the case of Gd 
isotopes, where they are lower. The agreement with experiment is 
comparable in the various calculations. The various calculations differ as much 
as one order of magnitude depending on the nucleus. Then, this is the expected 
spreading of the half-lives caused by different approaches and by uncertainties 
associated to the various approximations and choice of parameters. Certainly, 
it will be very interesting to extend the measurements of half-lives in this 
mass region.

The effect of the FF contributions is still controversial and seems to be very 
different in different mass regions. Whereas they represent minor effects in 
regions around $N=50$, their effect could be more important around $N=82$ and 
especially in $N=126$ \cite{fang13,borzov03,zhi13}. Several calculations include 
first-forbidden transitions in the rare-earth mass region 
\cite{moller3,mustonen16,marketin16}, but the results are in many cases completely 
at variance.

\begin{figure*}[htb]
\centering
\includegraphics[width=90mm]{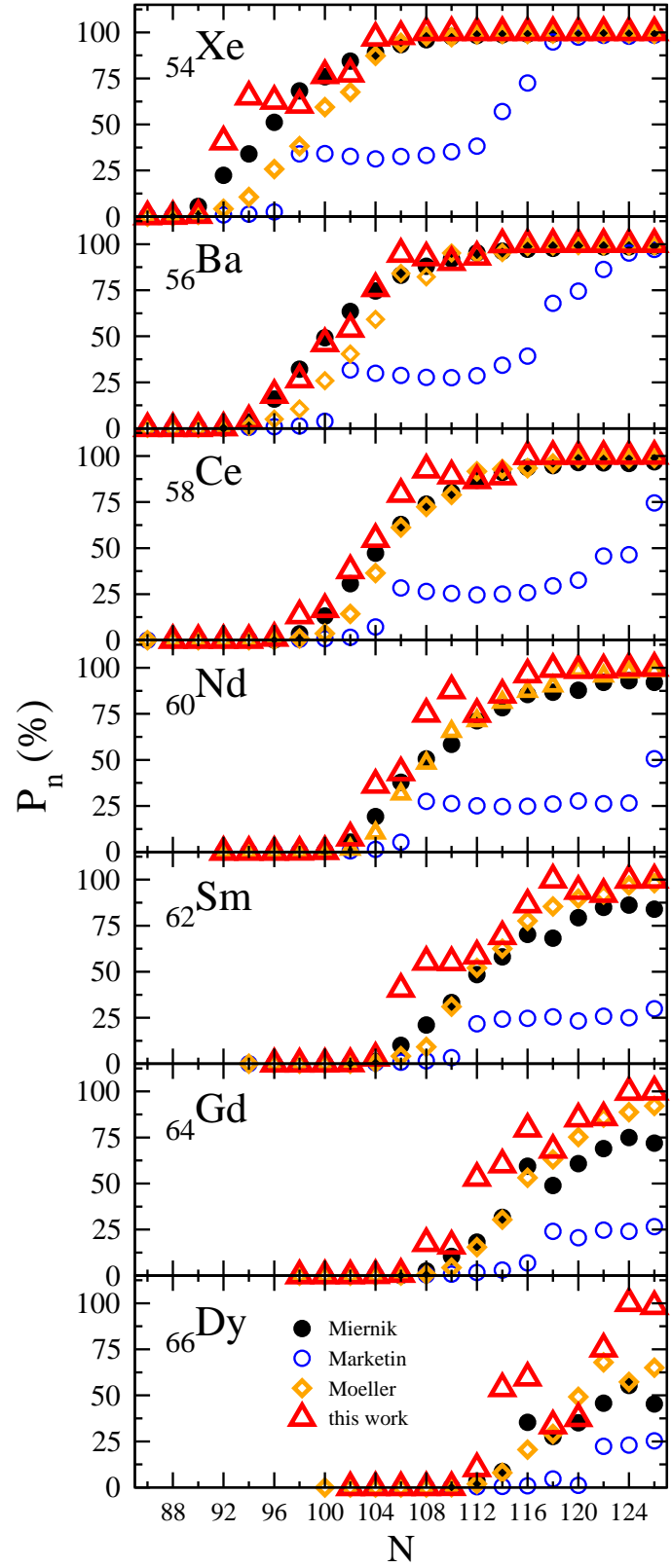}
\caption{Theoretical QRPA-SLy4 percentage values of the probability for $\beta$-delayed
neutron emission $P_n$. Results in this work are compared with those of 
Marketin {\it et al.} \cite{marketin16} and Moeller {\it et al.} \cite{moller3},
as well as with the phenomenlogical model by Miernik \cite{miernik}.
}
\label{fig_pn}
\end{figure*}

Figure \ref{fig_pn} shows the results for the $\beta$-delayed neutron-emission 
probabilities, expressed as percentages. These are the ratios of the 
rates due to transitions above the neutron separation energy $S_n$ to the total 
$\beta$-decay rates. As it is defined, $P_n$ includes the $\beta$-delayed 
probabilities for the emission of any number of neutrons. $P_n$ is a sensitive 
function of both $S_n$ and $Q_{\beta}$ energies, which are evaluated from the 
SLy4 Skyrme force \cite{masssly4}.

The lighter isotopes in all the isotopic chains in Fig. \ref{fig_pn} are close 
to stability and have $S_n$ energies larger than $Q_{\beta}$ energies. Therefore,
the $P_n$ values are obviously zero. As one moves away from stability, heavier 
isotopes exhibit decreasing $S_n$ energies and increasing $Q_{\beta}$ energies 
and as a result, $P_n$ values start to increase accordingly. The neutron number 
$N$ at which $P_n$ starts to increase corresponds to the isotopes where 
$S_n \simeq Q_{\beta}$. This neutron number increases with the number of protons 
and changes slightly depending on the model. In Fig. \ref{fig_pn} the results in 
this work are compared with both the results from relativistic calculations of 
Ref. \cite{marketin16} and from the microscopic-macroscopic calculations of 
Ref. \cite{moller3}. The increasing of $P_n$ with the nuclear instability is a 
general pattern in all the calculations, but whereas the results in this work 
follow the same tendency as the results in Ref \cite{moller3}, the $P_n$ values 
from the relativistic calculations are systematically lower. The direct 
consequence would be that less free neutrons are predicted in the astrophysical 
scenarios where these nuclei are decaying.
Unfortunately, the experimental information on $P_n$ values in this mass region
of interest for the {\it r} process is still very limited due to the low production
rates of exotic nuclei and to the difficulties inherent to neutron detection.
Indeed, there are no $P_n$  measurements available in the nuclei studied in this 
work. Thus, it is worth comparing the results obtained here with those of a
phenomenological model based on a statistical level density function. This 
model \cite{miernik} has been shown to reproduce the available experimental 
$\beta$-delayed neutron-emission probabilities to an equivalent or better degree 
than previous models.
The results shown in Fig. \ref{fig_pn} with black dots correspond to this model
\cite{miernik} based on theoretical masses from HFB-21 \cite{goriely10}
for the calculation of $Q_\beta$ and neutron separation energies. 
The isotopic pattern is in general fairly well reproduced, although the detailed
behavior in the region of increasing $P_n$ contains more fluctuations in the
present calculations.

\section{CONCLUSIONS}

I have calculated $\beta$-decay properties of even-even neutron-rich isotopes
in the rare-earth mass region. These nuclei are expected to play an important
role in the nucleosynthesis {\it r} process and in particular, they are crucial
to understand the existence of the rare-earth peak in the pattern of isotopic
abundances. Namely, isotopes of Xe, Ba, Ce, Nd, Sm, Gd, and Dy with neutron 
numbers between $N=86$ and $N=126$ are included in this study.

A theoretical approach based on deformed HF+BCS+QRPA calculations with Skyrme 
effective interactions and separable residual forces is used to obtain GT
strength distributions, $\beta$-decay half-lives, and $\beta$-delayed 
neutron-emission probabilities. The results from this approach are compared 
with the available experimental information and with calculations based on 
different methods. 
In general, a reasonable agreement with experiment is obtained. The results 
are comparable to other calculations using different approaches, using 
different mean fields or different residual interactions.

These calculations are timely because they address a mass region which is at 
the borderline of present experimental capabilities. Experimental information 
on the energy distribution of the GT strength is a valuable piece of knowledge 
about the nuclear structure in this mass region. The study of these distributions 
is within the current experimental capabilities in the case of the lighter 
isotopes considered in this work. Here, theoretical predictions have been 
presented for them based on microscopic calculations. Similarly, measuring the 
half-lives of the heavier isotopes will be highly beneficial to model the 
{\it r} process and to constrain theoretical nuclear models. These measurements 
are also a real possibility within present capabilities at MSU and RIKEN.

A data set containing the main results of this work is available as Supplemental 
Material \cite{suppl} to this article.

\begin{acknowledgments}

I thank K. Yoshida for useful discussions. This work was supported by Ministerio 
de Econom\'\i a y Competitividad (Spain) under Contract FIS2014-51971-P.

\end{acknowledgments}

\begin{figure*}[htb]
\centering
\includegraphics[width=120mm]{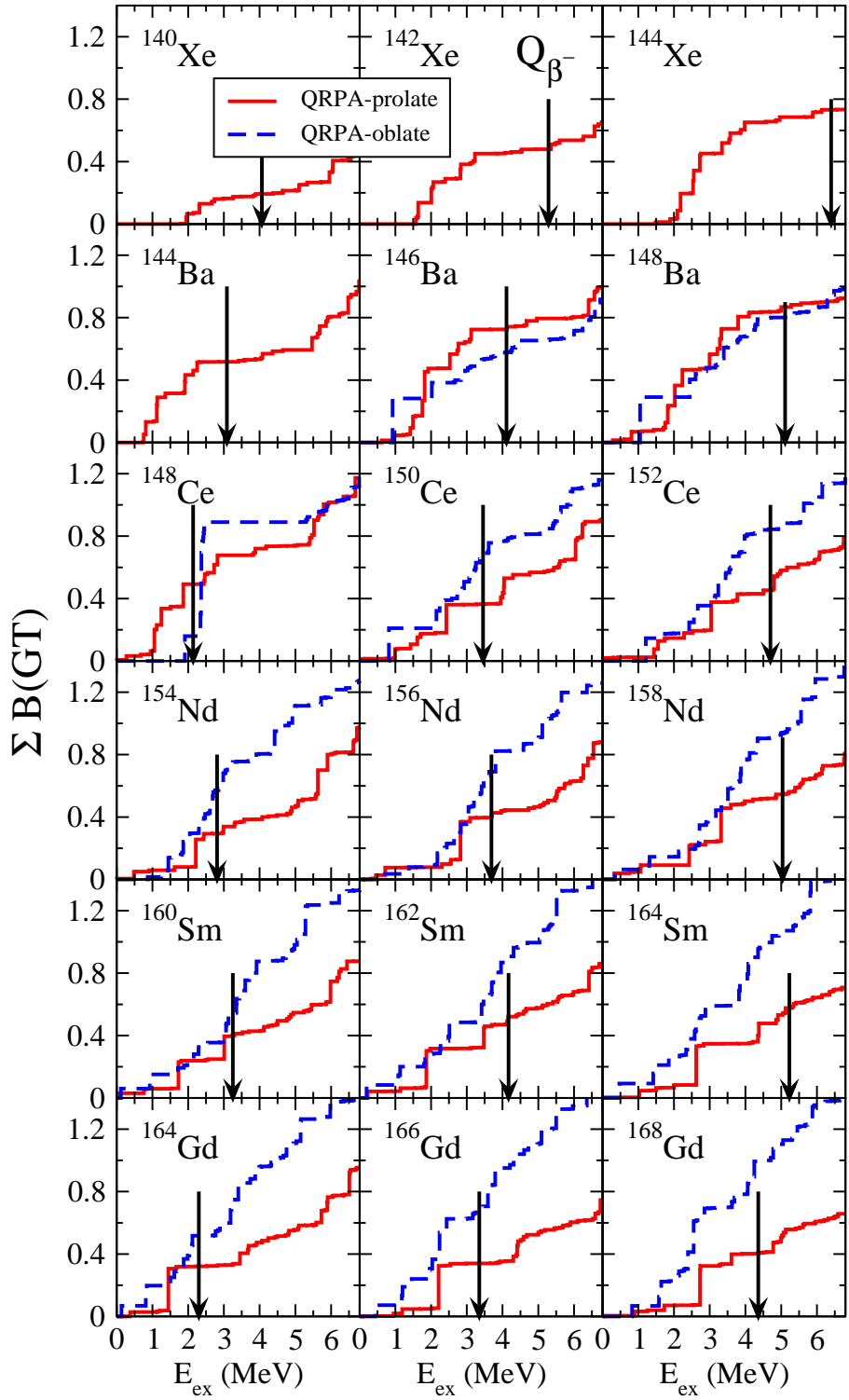}
\caption{(Color online) QRPA-SLy4 accumulated GT strengths in various Xe, Ba,
Ce, Nd, Sm, Gd, and Dy isotopes calculated for the prolate and oblate equilibrium 
shapes. $Q_\beta$ energies are shown with vertical arrows.}
\label{fig_bgt}
\end{figure*}

\end{document}